\documentclass[a4paper,11pt]{article}
\pdfoutput=1 
\usepackage{jinstpub} 

\usepackage{amsmath,amssymb}
\usepackage{amsfonts}
\usepackage{tabularx}
\usepackage{enumerate}
\usepackage{graphicx}
\usepackage{dcolumn}
\usepackage{url}
\usepackage{bm}
\usepackage{subfig}
\usepackage{float}
\usepackage{verbatim}
\usepackage[utf8]{inputenc}
\usepackage[english]{babel}
\usepackage{algorithm}
\usepackage{algpseudocode}
\usepackage{siunitx}
\usepackage[]{hyperref}

\makeatletter
\providecommand\add@text{}
\newcommand\tagaddtext[1]{%
  \gdef\add@text{#1\gdef\add@text{}}}%
\renewcommand\tagform@[1]{%
  \maketag@@@{\llap{\add@text\quad}(\ignorespaces#1\unskip\@@italiccorr)}%
}
\makeatother
\graphicspath{{Images/}}

\providecommand{\e}[1]{\ensuremath{\times 10^{#1}}}

\title{Estimation of cosmic-muon flux attenuation by Monserrate Hill in Bogota}

\author[1]{Juan Sebastián Useche Parra \note{Corresponding author.}}
\author[]{Carlos Arturo Ávila Bernal}

\affiliation[]{Physics Department, Universidad de Los Andes\\ Bogotá, Colombia}

\emailAdd{js.useche10@uniandes.edu.co}
\emailAdd{cavila@uniandes.edu.co}

\abstract{When designing an experimental setup for measuring the flux of cosmic ray muons through a geological structure, it is crucial to make detailed estimates about the expected muon flux attenuation by the structure studied. In this way, the coverage area of the detectors, the data acquisition system and the time length of data taking can be planned accordingly to the aimed accuracy of the measurements.  We present here a computational study of the cosmic muon flux at the location of Monserrate Hill in Bogota, Colombia, using the CORSIKA cosmic ray simulator. We use ArcGIS package to estimate the mean depth muons have to travel through the mountain to reach the detector. We consider two different uniform densities of rock for the mountain to make estimates of the muon flux. Through the interpolation of stopping power available data, one can then determine the attenuation as a function of the muon incident energy. Our final result provides an estimation for the muon flux attenuation as a function of the location of the mountain where the muons pass through. We also supply estimated values for the time needed to reach an accuracy of about 3$\%$ in the measurement of the muon flux through Monserrate Hill. Our strategy can be easily adapted for muon flux studies at any other location in the world.}


\begin{document}
\maketitle
\flushbottom

\section{\label{sec:Introduction} Introduction}
Cosmic rays, primarily composed of protons and Helium nucleus from different astrophysical sources, are continuously arriving at the Earth.  The most energetic particles can penetrate the geomagnetic shielding and interact with molecules in the atmosphere (mainly Nitrogen and Oxygen) producing particle cascades that continue propagating through the atmosphere until reaching the Earth's surface. The most abundant electrically charged component of the particle showers arriving at the terrestrial surface are muons, which interact with matter through similar physical processes as electrons, mainly deceleration by bremsstrahlung radiation when they are very energetic \cite{tanabashi2018review}. However, the interaction cross section by bremsstrahlung is inversely proportional to the square of the mass of the incident charged particle. Since muons are about 200 times more massive than electrons, then they can enter much deeper into matter before they get absorbed and the distances they can penetrate depend on their initial energy.  

Muon imaging is a technique that uses the high depth penetration property of cosmic-ray muons to study the density of objects with sizes around 1000 m or below \cite{nagamine2003introductory}. This non-invasive method is feasible to apply in the investigation of the internal structure of mountains, volcanoes, and high size physical structures, in a similar way as x-ray imaging is performed for small size objects, with the additional advantage that muons come from a natural source of radiation, always present at no cost. The shadow produced by the object in the measured muon flux, as compared to the muon flux at open sky, together with the depth profile that muons have to traverse throughout the object, are the main values needed to determine the average density of the object as a function of the transverse coordinates \cite{lesparre2012density}.

The first practical application of muons known today came about two decades after their discovery; it consisted of the measurement of muon flux attenuation by the overburden of a tunnel in Australia, performed in 1955 \cite{george1955cosmic}.  The first muon imaging attempt known is the work by Alvarez et al. in 1970 \cite{alvarez1970search}, who used spark chambers triggered by scintillation counters to measure the angular distribution of muons that penetrated the Chephren pyramid in Egypt, looking for hidden chambers. Even though no evidence of new chambers was found, this work provided the first experimental proof of the feasibility of muon imaging through the measurement of cosmic-muon tracks. The same principle applied 47 years later; with more robust and several detector types (Nuclear emulsion films, gas detectors, and scintillator hodoscopes) allowed the discovery of a void, consistent with a hidden chamber, above the known Grand Gallery, in the Khufu's great pyramid in Egypt \cite{morishima2017discovery}.

Muon imaging has become an active area of research during the last two decades. Especially, as a result of its applications in Earth sciences estimating the internal structure of mountains and volcanoes.  In 1995 Nagamine et al. published their measurements of the inner structure of mount Tsukuba in Japan with the use of cosmic muons detected by a three-fold plastic scintillator telescope positioned at 2.0 km from the mountain \cite{nagamine1995method}.   Tanaka et al., have produced muon absorption images of the lava conduits of the Usu and Asama volcanoes in Japan \cite{tanaka2007high,tanaka2009detecting,tanaka2008radiographic}, and more recently they studied the dynamics of magma discharges in the active Satsuma-Iwojima volcano, also at Japan \cite{tanaka2014radiographic}. Cosmic muon tracks were also used to study the lava dome of La Soufriere volcano in France in 2012 \cite{lesparre2012density}. Three data-taking campaigns, between 2013 and 2015,  have been carried out by the TOMUVOL collaboration to obtain muon radiographic images of the  Puy de Dôme volcano in France, by the use of resistive plate chamber detectors \cite{le2016rpc}. More recently, the MU-RAY collaboration reported the use of cosmic-ray muons for searching underground cavities on mount Echia in Naples, Italy \cite{saracino2017imaging}. Muon imaging has become such a powerful tool for the exploration of geological structures that ideas of using it for the exploration of Mars' geology have already been proposed \cite{kedar2013muon}.

Additionally to the use of muon attenuation for imaging of high dense objects, new applications exist today which are based on measuring the deflection of the muon trajectory due to multiple scattering when traversing a material \cite{procureur2018muon}. The main constraint of this technique is the use of two detectors with high spatial resolution, one on each side of the object, to measure the trajectories of muons. On the other hand, the muon deviation method requires less time to reconstruct the studied object than absorption muography, but it only works for medium opacity materials \cite{procureur2018muon}. Borozdin et al., measured the scattering of muons due to the presence of a cylinder of tungsten and the reconstruction of its radiographic image \cite{borozdin2003surveillance}. These results show that one of the main applications of deviation muography is the imaging of medium-sized objects and the high contrast detection between materials of different densities.

The aim of our work is to propose a rather simple simulation method to estimate the attenuation of cosmic ray muon flux when traversing a mountain. We specifically work on the case of Monserrate Hill in Bogota and provide estimates of the muon flux rate for the tropical atmosphere, at an altitude near 2700 m above sea level, as well as estimates of muon flux attenuation. We start with a description of cosmic ray primary particles and their products (sections \ref{sec:PrimaryFlux} and \ref{sec:CRmuons}), then some general and interesting aspects of the Monserrate Hill are shown in section \ref{sec:Monserrate}. We describe in section \ref{sec:MuonFlux} the muon flux simulation together with the experimental validation we have performed. In section \ref{sec:MuonsMonserrate}, we present the results of muon transmission through Monserrate Hill and find the time to achieve about 3\% uncertainty in the determination of muon flux attenuation. Section \ref{sec:Conclusions} summarizes our conclusions.     
\section{\label{sec:PrimaryFlux}Primary Cosmic Rays}
Cosmic rays (CR) are one of the oldest sources of high energy particles that have been studied by mankind. When a primary cosmic-ray interacts with a molecule in the atmosphere it generates hadronic and electromagnetic showers \cite{tanabashi2018review}. The hadronic showers are mainly composed of pions, kaons and nucleons, while the electromagnetic part is made out of muons, electrons, positrons and photons. The later ones, are produced by the decay of neutral pions or by bremsstrahlung radiation from electrons and positrons. Muons are mainly produced by the decay of charged pions (eq. (\ref{eq:decaimientosHadronicos1})), which additionally generate electrons and neutrinos from their decay, contributing to the development of the electromagnetic cascade \cite{Grieder2001}.
\begin{equation}
  		\pi^+ \rightarrow \mu ^+ + \nu_\mu, \qquad
		\pi^- \rightarrow \mu ^- + \bar{\nu}_\mu
		\label{eq:decaimientosHadronicos1}  
\end{equation}

One of the main aspects of primary cosmic rays is their chemical composition, that can be studied using the energy spectrum and abundance of each chemical compound as in \cite{Wiebel-Sooth1998}. The flux ($j$) as a function of the energy ($E$) of these primary particles can be approximated as a power law \cite{Grieder2001} in the form:

\begin{equation}
    j(E)=j_0E^{\alpha}
\end{equation}

\begin{figure}[H]
    \centering
    \includegraphics[width=0.8\textwidth]{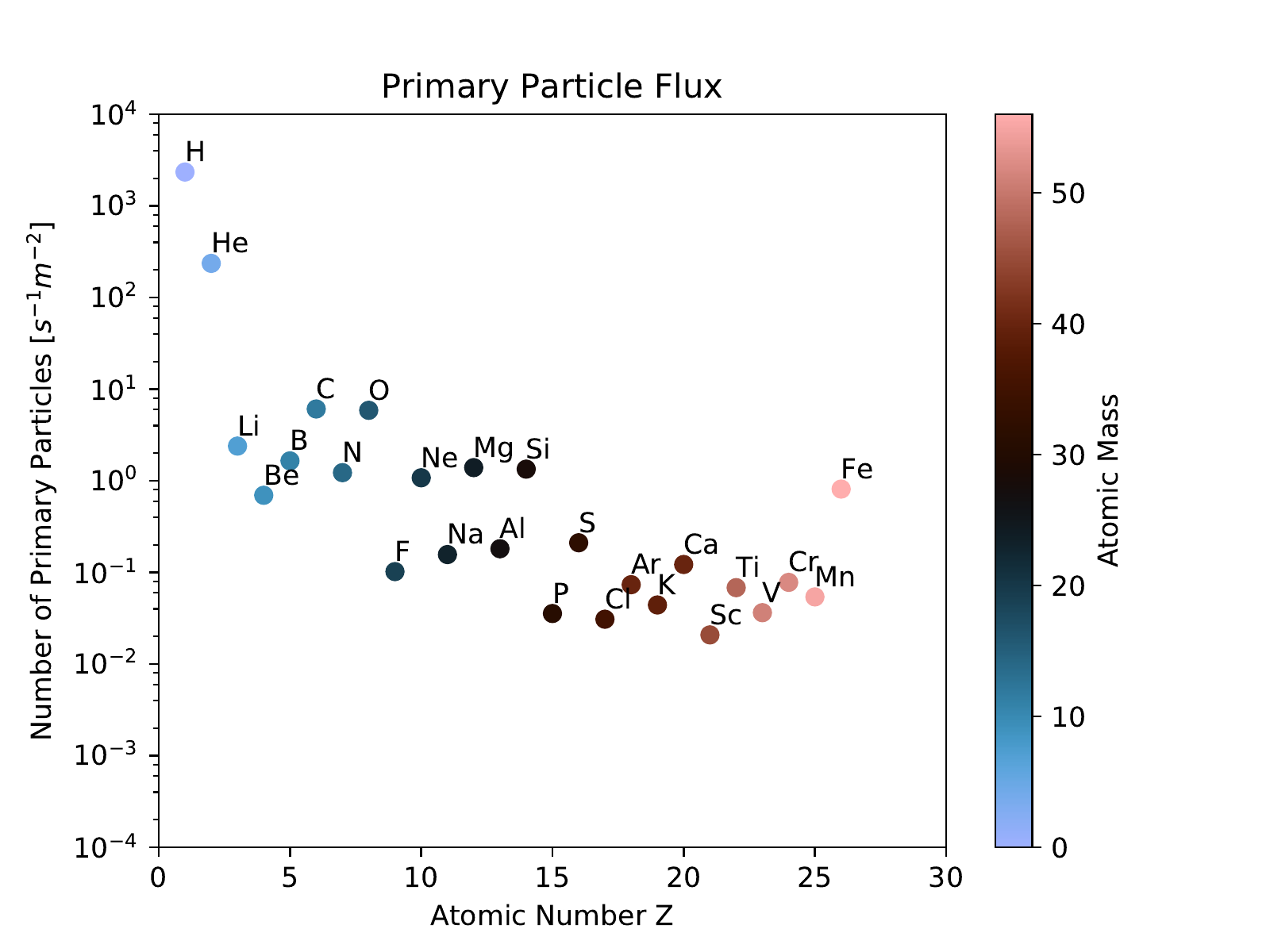}
    \caption{Flux of primary particles calculated using eq. (\ref{eq:PrimaryFlux}) and the data from \cite{Wiebel-Sooth1998}.}
    \label{fig:PrimaryFlux}
\end{figure}

With $\alpha < 0$, meaning a decrease of particles with energy. The flux corresponds to the ratio of the number of particles over $area \times time \times solid$ $angle \times energy$. Therefore, the power law can be rewritten as:
\begin{equation}
    \frac{dN(E)}{dAd\Omega dtdE}=j_0E^{\alpha}
\end{equation}
Assuming that the number of particles is constant over the area \textit{A} of the flat detector and integrating over time, solid angle and energy, the primary particle flux is given by:
\begin{equation}
    \frac{N(E)}{A\Delta t}=\pi j_0 \sin^2(\theta_{max})\frac{E^{\alpha+1}}{\alpha+1}
    \label{eq:PrimaryFlux}
\end{equation}
Where $\theta_{max}$ is the maximum zenith angle that a primary will have when entering to the atmosphere. Figure \ref{fig:PrimaryFlux} displays the results of solving eq.(\ref{eq:PrimaryFlux}) for each element with atomic number Z, using the energy range  $Z\times10$ $GeV\leq E \leq 10^{6}$ $GeV$. The lower limit corresponds to the minimum magnetic rigidity ($R=\frac{E}{Ze}=10$ $GV$) for primary particles to have a negligible deflection by the Earth's magnetic field \cite{tanabashi2018review},\cite{Asorey2012}.

The Earth's geomagnetic field plays an important role in the number of primary particles that arrive and interact in the atmosphere. The main effect of the geomagnetic field is the shielding or cutoff in energy for a given particle to enter into the atmosphere. This cutoff is given by three main effects:
\begin{itemize}
    \item \textbf{Latitude effect:} The CR intensity shows a variation with latitude for energies less than 15 GeV, proved by A.H. Compton in 1936 \cite{Grieder2001}. This effect is also associated with the change in temperature of the atmosphere, for example, one-third of the CR intensity difference at sea level has been confirmed to come from a rise in the atmosphere temperature \cite{rossi1964cosmic}.
    \item \textbf{Longitude effect:} The variation of CR intensity with the longitude is due to the asymmetry of the geomagnetic dipole axis with respect to the Earth's rotation axis \cite{Grieder2001}.
    \item \textbf{East-west asymmetry:} The flux coming from the east and west direction presents a difference in energy (intensity) up to 100 GeV. This difference is more noticeable at higher altitudes (top of the atmosphere) than at sea level, because the flux suffers a modulation due to the dependence on the zenith angle (going as $\propto cos^2(\theta)$) that screens this difference of flux at sea level \cite{Grieder2001}. This effect was first measured by three experiments at distinct locations, reporting differences from 10\%-26\% between the east-west cosmic-ray fluxes, concluding that a higher fraction of the flux was positively charged particles \cite{rossi1964cosmic}.
\end{itemize}

\section{\label{sec:CRmuons} Cosmic ray muons}

The behavior of CR muons has been broadly studied, one of the main properties is their intensity following a cosine distribution of the zenith angle, $\theta$, \cite{Grieder2001} in the way described by:

\begin{equation}
I(\theta) = I_0 \cos^n(\theta) 
\end{equation}

Experimentally the parameter \textit{n} has been measured to be \cite{Grieder2001} \cite{Judge1965}:

\begin{equation}
    n=1.96\pm0.22
\end{equation}

Also, the experiment of Morris et al. \cite{Morris2014} proves that CR data taken at an altitude of about 2 km is consistent with an exponent parameter of $n=2$. On the other hand, one must expect a mostly uniform distribution along the azimuth angle, but due to the presence of Earth's magnetic field and the stochastic behavior of the cosmic showers, some azimuth fluctuations can be present at a low-level \cite{Grieder2001}.

\subsection{\label{subsec:Interaction} Muon-Matter Interaction}

Any charged particle interacts with matter through different processes such as radiation, pair production, bremsstrahlung, and photonuclear effects \cite{tanabashi2018review}. In the case of CR muons, this interaction is lead by radiation processes and the mean energy loss, with respect to the mass per unit area ($x$) of the material, in units of $MeV g^{-1}cm^2$ can be described using the following equation \cite{tanabashi2018review}:

\begin{equation}
    \Big<\frac{-dE}{dx}\Big>=a(E)+b(E)E 
    \label{eq:dEdx}
\end{equation}   

Where $b(E)$ corresponds to the sum of pair production, bremsstrahlung and photonuclear contributions and $a(E)$ accounts for ionization losses and is given by the Bethe-Bloch formula:

\begin{equation}
 	\Big<-\frac{dE}{dx}\Big>=Kz^2\frac{Z}{A}\frac{1}{\beta}\Big[\frac{1}{2}\ln\frac{2m_ec^2\beta^2\gamma^2T_{max}}{I^2}-\beta^2-\frac{\delta}{2}\Big]
    \label{eq:bethe}   
\end{equation}

 Where  $Z $ and $A$ are the atomic and mass numbers of the absorber respectively; $K$ is a constant with approximate value of $0.307$ $MeV cm^{2}mol^{-1}$, $z$ is the charge of incident particle, $T_{max}$ corresponds to the maximum energy received by an electron through a collision, $I$ is the mean excitation energy and $\delta$ is the density effect correction to ionization energy loss, $\beta$ and $\gamma$ correspond to the ratio of the particle velocity over $c$ and the Lorentz factor respectively \cite{tanabashi2018review}. Eq.(\ref{eq:bethe}) represents the mean rate of energy loss by relativistic heavy charged particles due to ionization, it is valid in the range $0\lesssim \beta\gamma\lesssim  1000$ \cite{tanabashi2018review}.

\begin{figure}[ht]
    \centering
    \includegraphics[width=0.8\textwidth]{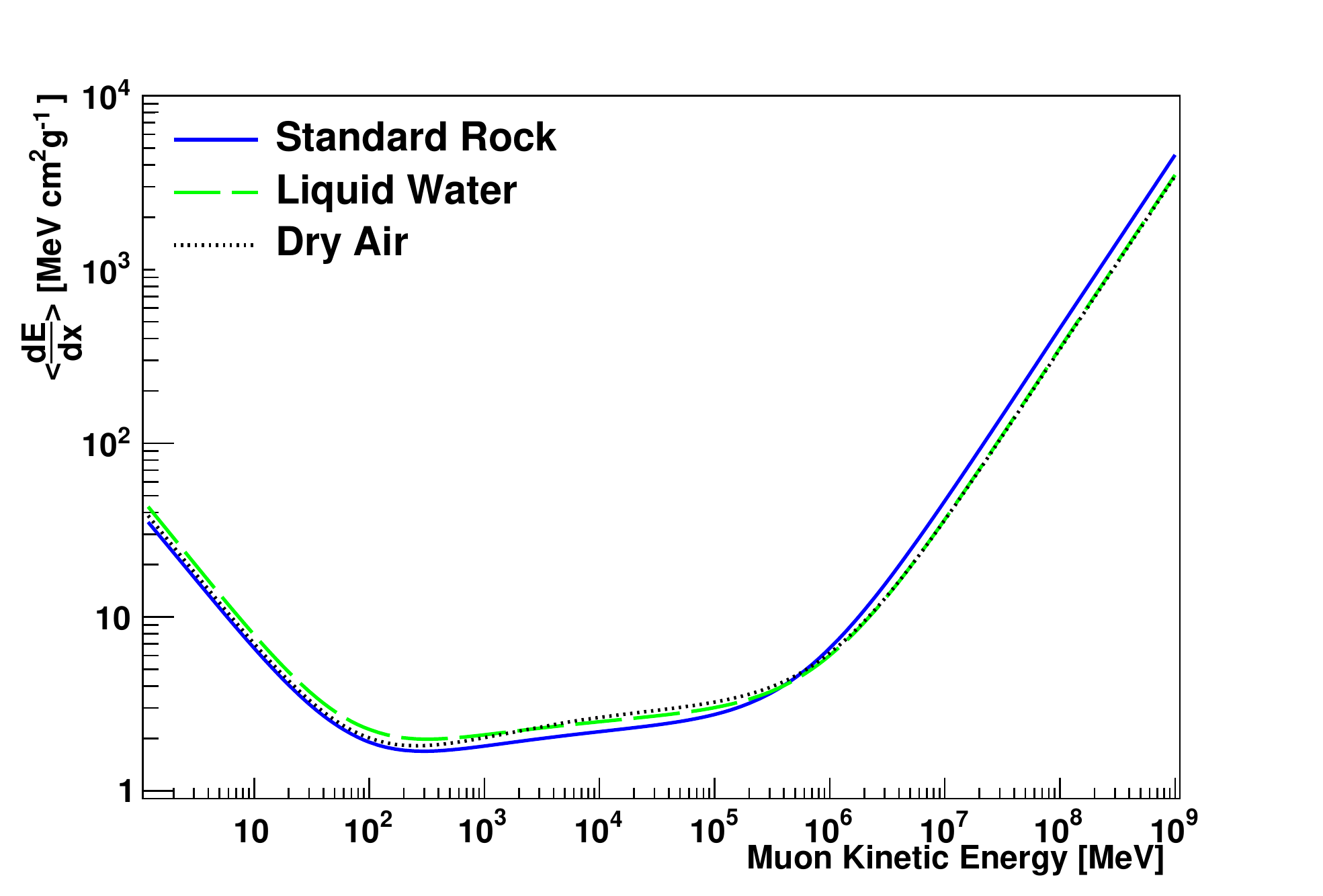}
    \caption{Energy loss of muons for different materials. The plotted data is taken from the Particle Data Group tables in \cite{tanabashi2018review}.}
    \label{fig:dEdxMuons}
\end{figure}

At higher energies, the energy loss of muons is led by radiation processes, then the  $b(E)E$ factor in eq.(\ref{eq:dEdx}) becomes important. The energy boundary where ionization losses and radiative effects are comparable is defined as the muon critical energy ($E_{\mu c}$) in the material \cite{tanabashi2018review}. For standard rock this critical energy is $E_{\mu c}\approx 6.9\e{2}\;GeV$ \cite{tanabashi2018review}. In figure \ref{fig:dEdxMuons} the nonlinear behaviour of energy loss for muons crossing trough standard rock ($\rho = 2.65\;gcm^{-3}$), liquid water ($\rho = 1.00\;gcm^{-3}$) and dry air ($\rho = 1.20 \e{-3}\; gcm^{-3}$) is plotted as a function of muon incident energy.

The quantitative study of the passage of muons through a material by the use of eq.(\ref{eq:dEdx}) \cite{GROOM2001} may require a large computation time. An alternative to this problem is to interpolate the energy loss due to the passage of muons inside a small portion of the material and keep recalculating it through the whole length needed, using the incidence energy and the energy loss per $cm^2/g$ of the material, tabulated by the Particle Data Group (PDG) \cite{tanabashi2018review}, as explained in section \ref{sec:MuonsMonserrate}. This last approach is used by us iteratively to calculate the energy loss of muons entering to standard rock with a given initial energy and a 1 m step of interaction inside the material.

\begin{figure}[ht]
    \centering
    \includegraphics[width=0.8\textwidth]{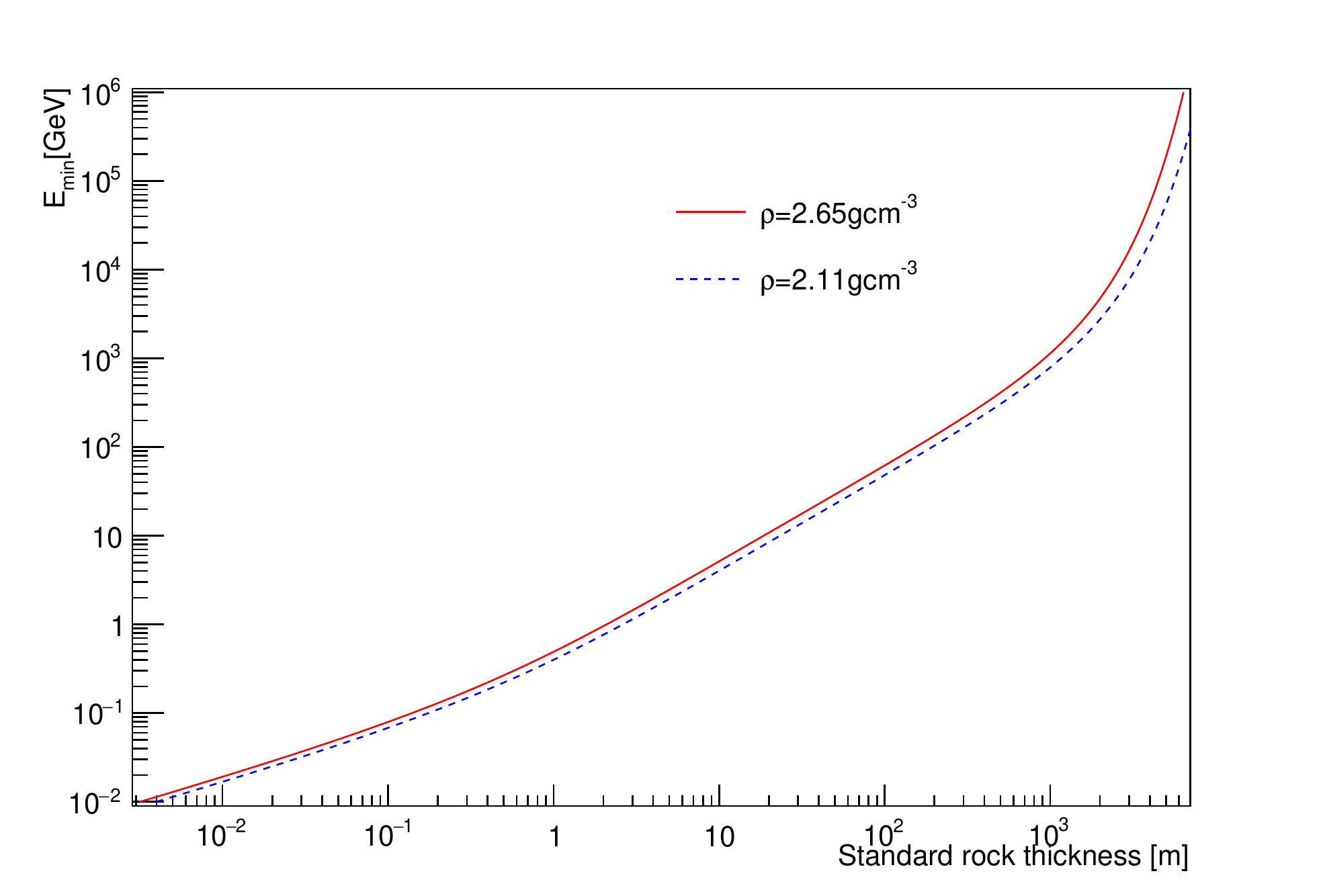}
    \caption{Minimum energy as a function of rock thickness and density of rock, based on the data provided by \cite{tanabashi2018review}.}
    \label{fig:EminCSDA}
\end{figure}

Also, it is possible to use the Continuous slowing down approximation (CSDA) range summarized in \cite{tanabashi2018review} and estimate the relationship between the minimum energy of a muon ($E_{min}$) needed to cross a portion of standard rock, as shown in figure \ref{fig:EminCSDA} for two rock density values. Most of the rock types have, in average, the same atomic number to mass number ratio $(\langle Z/A \rangle=0.5) $, with variation in density \cite{lesparre2010geophysical}, then it is possible to use the same parameters ($a(E)$, $b(E)$) to calculate the minimum energy to cross a portion of a rock-type material, with the ones given for standard rock in \cite{tanabashi2018review}.

\section{\label{sec:Monserrate} General aspects of Monserrate Hill}

Monserrate Hill corresponds to the western edge of Bogota mountainous area, originated by Bogota's fault. This fault is primarily composed of layers of clay and sandstone of different sizes, also occasionally there are some layers of phosphorite \cite{Uscategui2005}. This Hill was chosen for this study, due to the proximity to our university campus ($\sim1.2\;km$), also there are very few studies \cite{lobo1992geologia} about this Hill and it is important to expand the knowledge about it. It is located towards the east of downtown Bogota. It's highest point is  at 3176 meters above mean sea level (masl) with GPS coordinates  $4^\circ36'18''\;N$ $74^\circ03'19''\;W$. The Hill is an important touristic place in the city, that's why there are several locations on it that can be used as a reference for defining muon orientation, these reference sites are shown in Table \ref{table:Monserrate}. \\

\begin{table}[H]
	\centering
	\caption{GPS localization for reference points in Monserrate Hill. Taken from \cite{Guerrero2016}.}
	\begin{tabular}{p{0.14\textwidth}|p{0.14\textwidth}|p{0.11\textwidth}} 
		\hline
		\hline
Place & Location & Height [m]\\
		\hline
		\hline
Monserrate Church &$4^\circ36.315'N$ $74^\circ3.330' W$ & 3189\\
		&&\\
Funicular base entrance &$4^\circ36.174'N$ $74^\circ3.672' W$ & 2694\\
        &&\\
Funicular top entrance & $4^\circ36.213'N$ $74^\circ3.310' W$ & 3144\\
    \hline
    \hline
\end{tabular}
\label{table:Monserrate}
\end{table}

Using ArcGIS framework (Version 10.4) \cite{Esri}, which is a geographic information system to gather, analyze and visualize data using maps; and a Digital Elevation Model (DEM), with a 10m spatial resolution, provided by NASA \cite{NASA}, it is possible to calculate the depth of the mountain in a given direction. Eleven profiles for the azimuth angle (with a $\Delta \phi = 5^\circ$ separation) were taken, using the church of Monserrate as a reference for $\phi=0^\circ$. For the zenith angle, we also use a separation of $\Delta \theta =5^\circ$. With this angular coverage, we end up with cells of $105\;m\times105\;m$ projected onto the mountain's surface, as seen from a point in the university campus at an approximate distance of $R\approx1.2\;km$.

\begin{figure}[ht]
    \centering
    \includegraphics[width=0.8\textwidth]{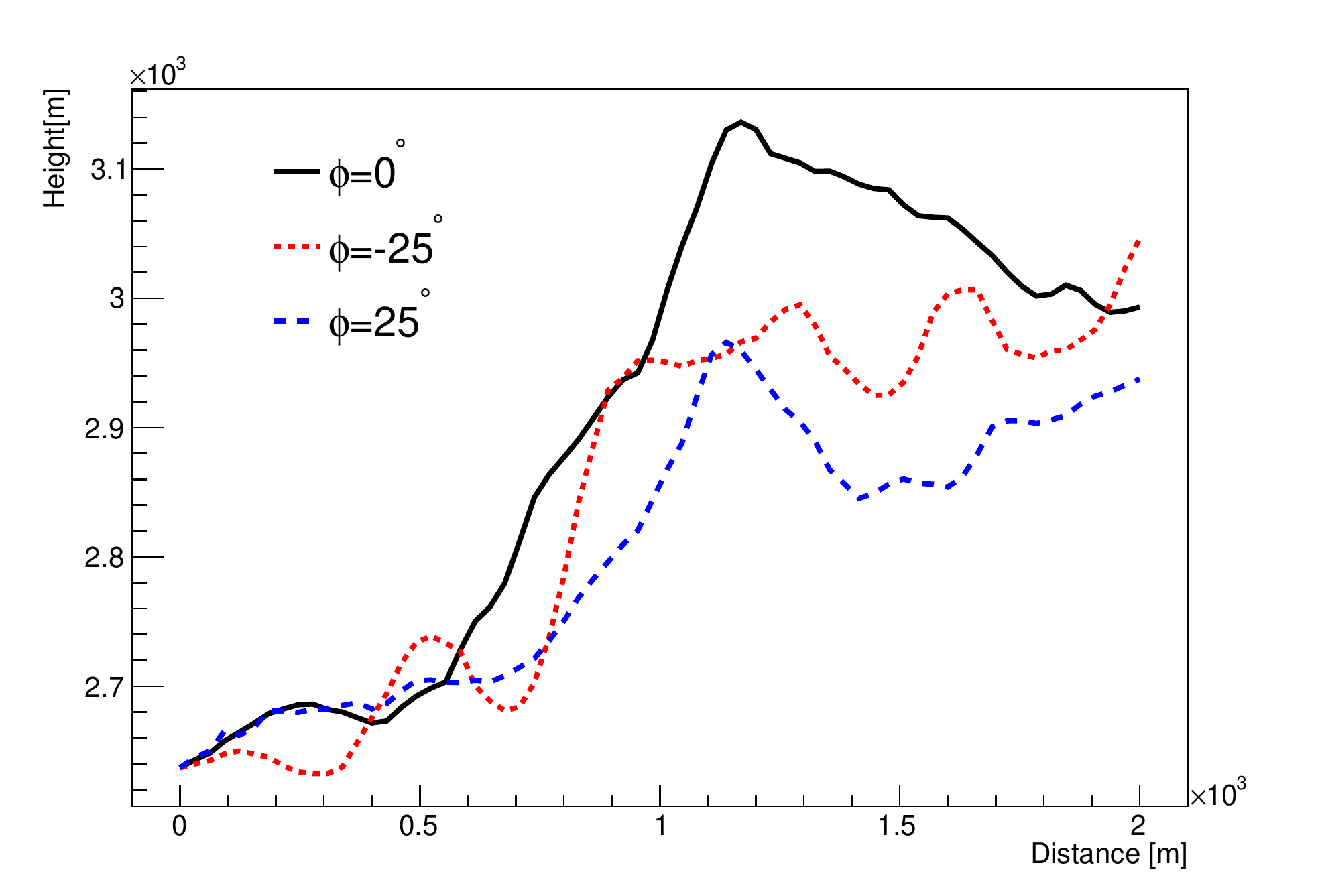}
    \caption{Elevation vs transverse length profile of Monserrate Hill in meters, using ArcGIS software \cite{Esri}. Only profiles for $\phi = 0^\circ,-25^\circ,25^\circ$ are displayed.}
    \label{fig:ProfilesMonserrate}
\end{figure}

\begin{figure}[ht]
    \centering
    \includegraphics[width=0.8\textwidth]{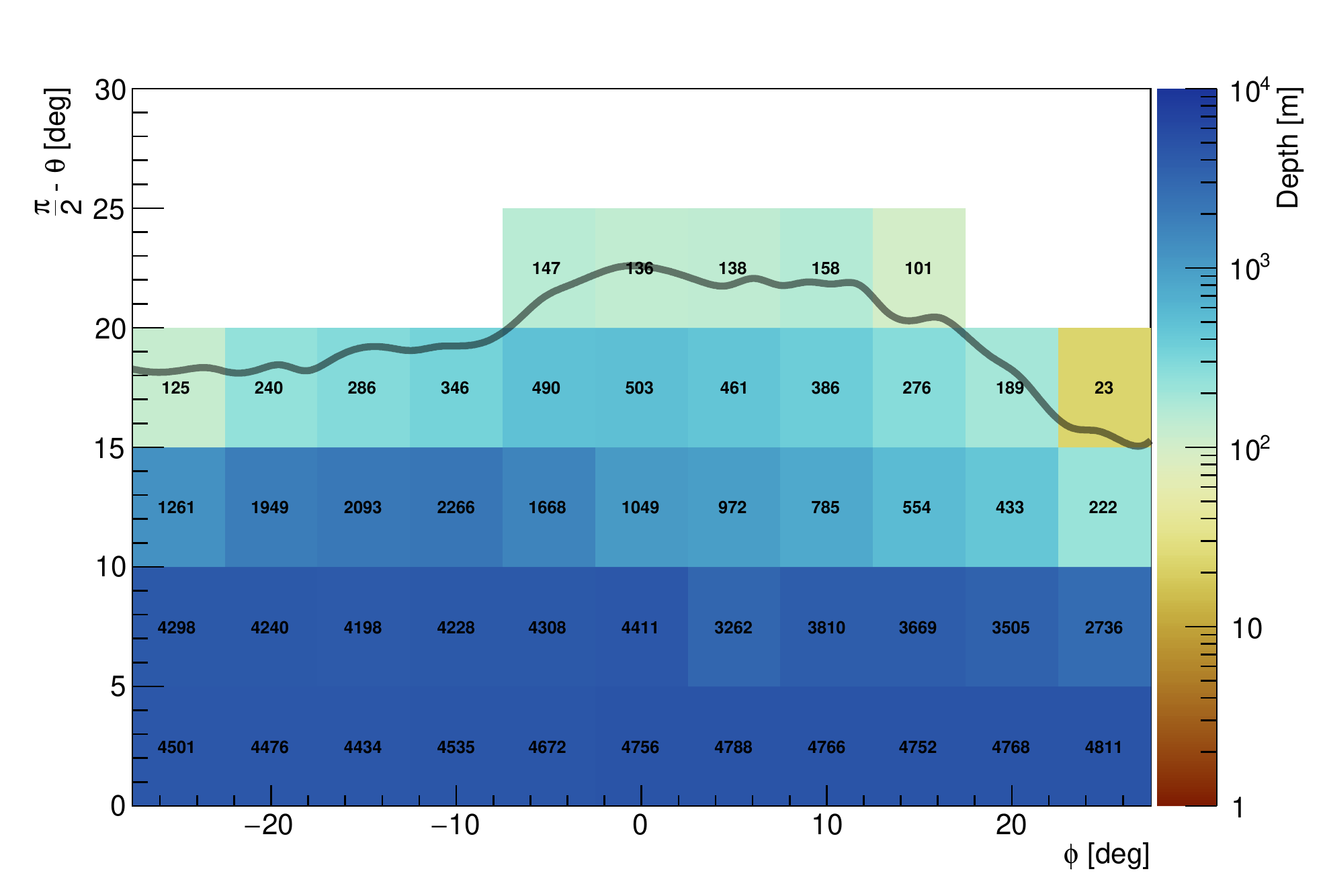}
    \caption{Mean depth of Monserrate Hill in meters, for different cells of the mountain as a function of $(\theta,\phi)$. The black solid line corresponds to the transverse border of the mountain as seen from the point of observation in the university campus.}
    \label{fig:LengthMonserrate}
\end{figure}

\begin{figure}[ht]
    \centering
    \includegraphics[width=0.8\textwidth]{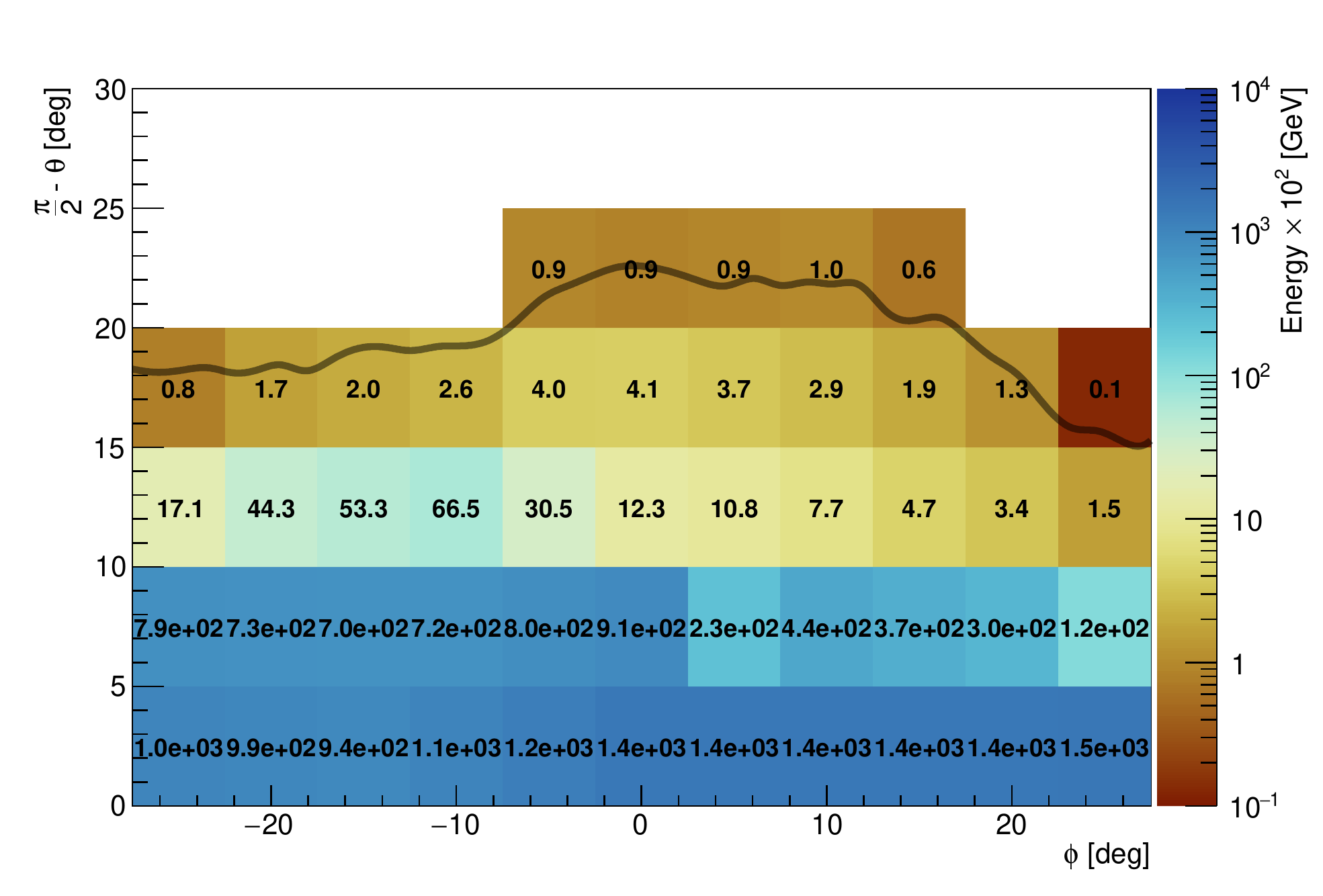}
    \caption{Estimation of minimum energy needed for a muon to cross Monserrate Hill, assuming a uniform density of $\rho=2.65\;gcm^{-3}$ (standard rock).}
    \label{fig:EminMonserrate}
\end{figure}

In this work, the viewpoint was chosen as the top of a building within the university campus with latitude $4^\circ36'10''\;N$, longitude $74^\circ 3' 53''\;W$ and an altitude of 2660 masl. The elevation profile of the mountain, obtained with ArcGIS, is shown in figure \ref{fig:ProfilesMonserrate}. Note that the elevation is measured from the floor of the building, 20 m below our observation point.
For the estimation of the muon flux attenuation, we use the mean depth penetration on a given cell (as reported in figure \ref{fig:LengthMonserrate}), obtained by dividing the cell in intervals of $\Delta \theta=1^\circ$. By performing a linear interpolation of the two closest CSDA values to this length, it is possible to estimate the minimum muon energy needed to cross the mountain's cell, the results of the interpolation are shown in figure \ref{fig:EminMonserrate}.

\section{\label{sec:MuonFlux} Muon Flux Simulation}
The simulation work was performed in two stages, the first involved the simulation of cosmic ray showers using the software CORSIKA-v75600 \cite{Heck2016} with the ROOTOUT option to store the results in a ROOT \cite{brun1997root} file. Each simulated shower takes into account the number of primary particles per second per area that produce a shower according to their chemical composition, using the results of eq.(\ref{eq:PrimaryFlux}) in figure \ref{fig:PrimaryFlux}. Then, each primary is given a direction of incidence by a Monte Carlo (MC) process in CORSIKA, within the intervals $0^\circ\leq \theta \leq 88^\circ $  and $-180^\circ \leq \phi \leq 180^\circ$. Finally, the location's altitude, atmospheric model and values for the vertical and horizontal geomagnetic field, which can be calculated using the program \textit{Geomag} developed by the National Oceanic and Atmospheric Administration (NOAA) and available on-line in \cite{Geomag} , must be introduced as input to CORSIKA.

In our case, as multiple showers are simulated at once, the single-core feature of CORSIKA is used. This imposes some restrictions in the simulation architecture that are solved by using a high-processing computer and executing in multiple nodes at the same time. The configuration of our system was one node with 3 processors and RAM memory of 26 GB, achieving a computational time of 4 hours of real time per 2 hours of simulation.

The second step is the extraction of muon incident angle and momentum in the detector, which are extracted from CORSIKA output files. With this information, we reconstruct the energy and angular distribution of muons in a certain location and simulate the passage of muons throughout the mountain.

\subsection{\label{subsec:FluxMeasurement} Near Horizontal muon flux verification}
The first verification of the simulation method is to replicate experimental data at sea level. We use data from the muon flux spectrum measurements performed by the DEIS spectrometer of the Kiel-Tel Aviv collaboration \cite{allkofer1985cosmic}, to compare to our CORSIKA simulation. These data are also used by other authors to compare their  cosmic ray simulators \cite{nishiyama2016monte}. The parameters used for the simulation are an altitude of 5 masl and GPS location of $32^\circ6'42.4''N$, $34^\circ48'4.9''E$, which gives a geomagnetic field of $B_x=29.9\;\mu T$ and $B_z=33.5\;\mu T$.

\begin{figure}[ht]
    \centering
    \includegraphics[width=0.8\textwidth]{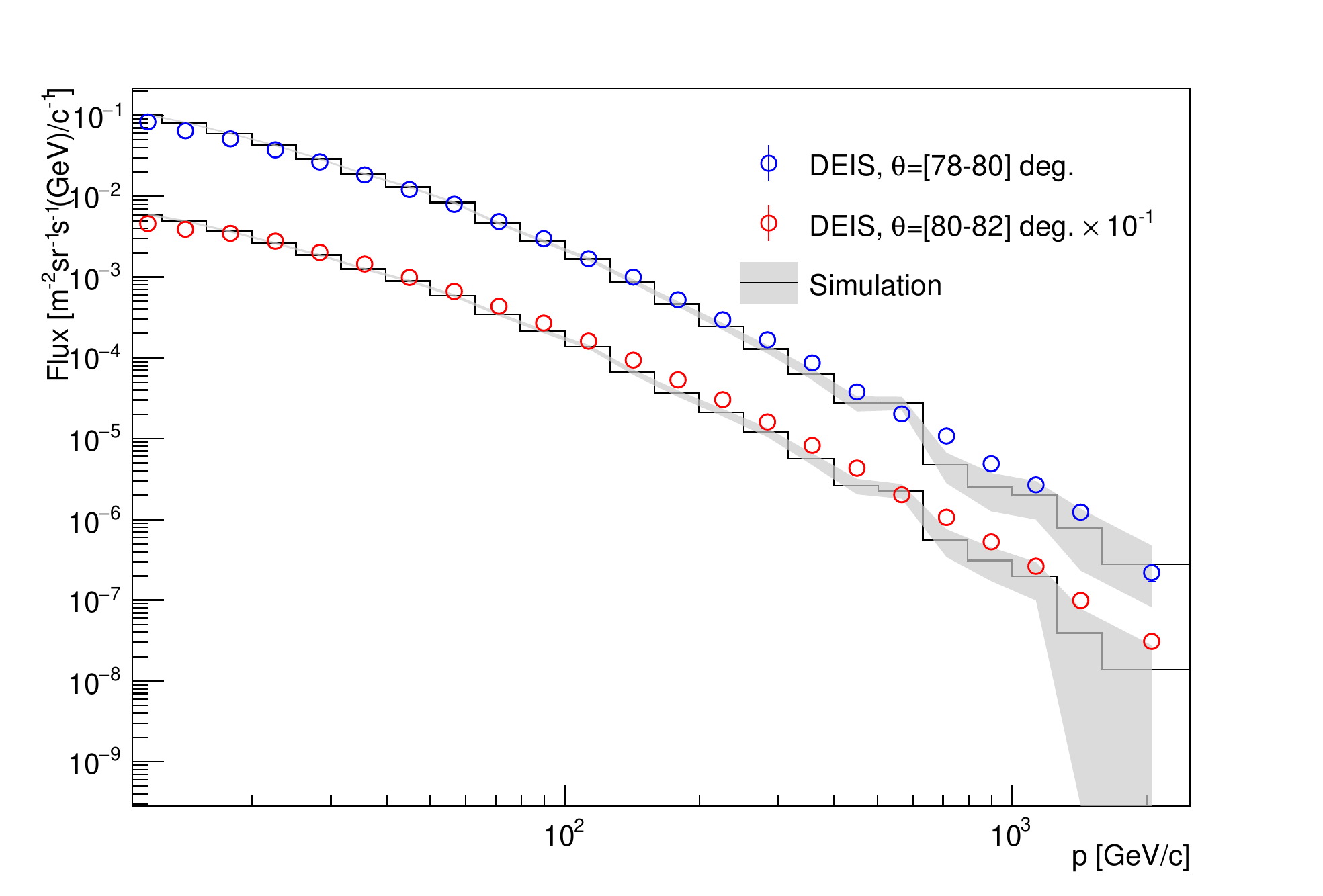}
    \caption{Muon energy spectra from a simulation of 10 hours of cosmic ray flux arriving at Tel Aviv, Israel, compared to DEIS experimental data \cite{allkofer1985cosmic}. The simulated spectra is compared in the intervals of zenith angles $\theta=[78:80]$ and $\theta=[80:82]$. The indicated one standard deviation error band is due to the statistical uncertainty of the simulation.}
    \label{fig:MomentumComp}
\end{figure}

The DEIS experiment gathered data for zenith angles in the region $\theta=[78^\circ:90^\circ]$. When comparing the integral of the energy spectra distributions from data and simulation, we find that CORSIKA is simulating in average 20\% fewer muons than experimentally reported. This level of agreement is better than the one achieved by the simulation reported in \cite{nishiyama2016monte}. This discrepancy is mainly due to the atmospheric model used by us, which is \textit{T616 Central European atmosphere}, given that the experimental measurements are very sensitive to atmospheric variations. 

\subsection{\label{subsec:FluxMeasurement2} Mid-range flux verification}

We compare our simulation to data gathered by the Okayama cosmic-ray telescope \cite{Tsuji1998}. For this purpose, cosmic ray muons were simulated for a zenith angle of $\theta=30^\circ \pm 4^\circ$ and azimuth angle of $\phi= 130^\circ \pm 7.75^\circ$ (north-west direction), an altitude of 5.3 masl with GPS coordinates $34^\circ N$, $133^\circ56' E$, the \textit{mid-latitude summer} model for the atmosphere from the Bernlohr package \cite{Heck2016}, included in CORSIKA libraries, was used together with values of $B_x=31.3\;\mu T$, $B_z=35.4\;\mu T$ for the geomagnetic field.

\begin{figure}[ht]
    \centering
    \includegraphics[width=0.8\textwidth]{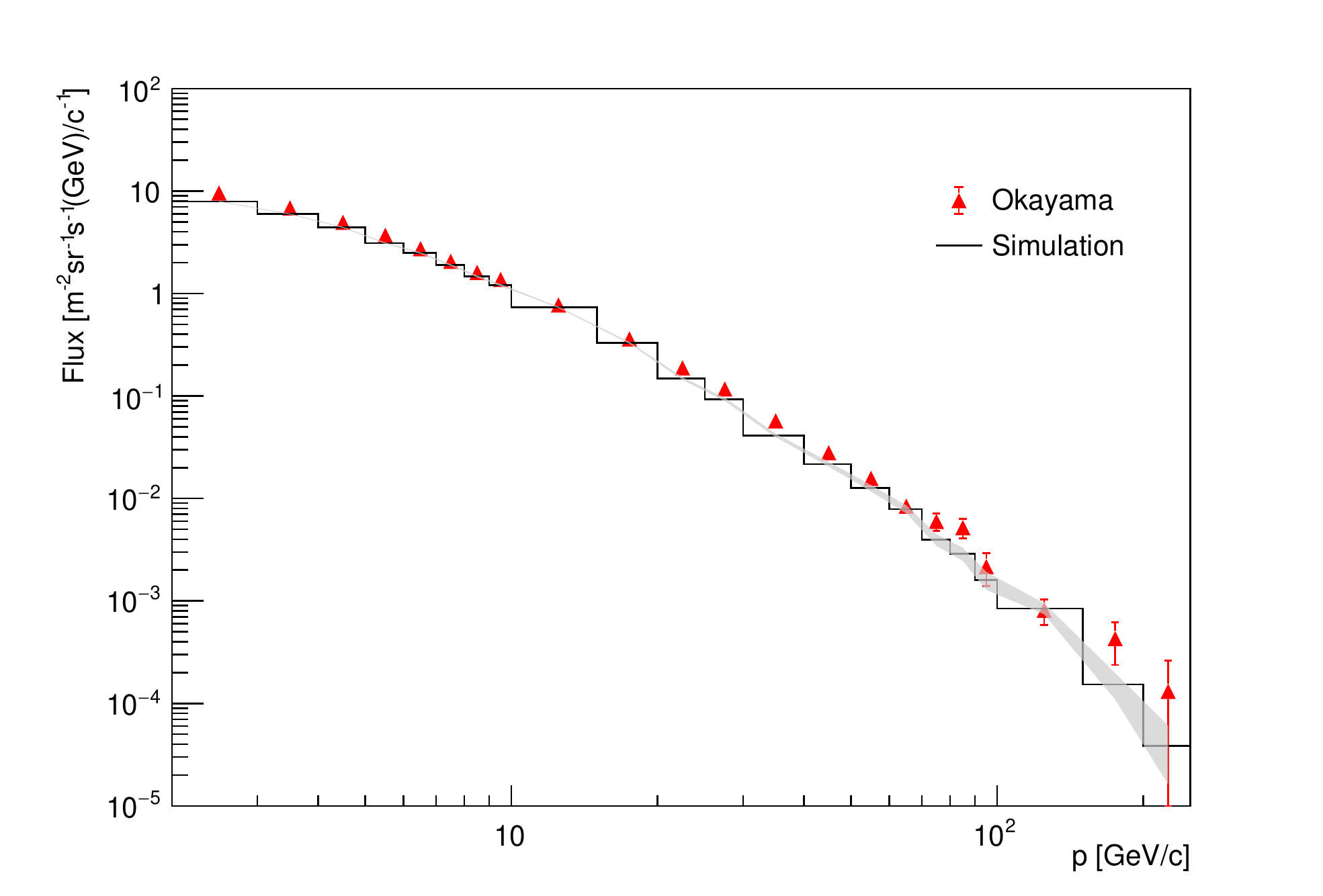}
    \caption{Muon spectra at sea level for an azimuth angle of $\theta=30^\circ \pm 4^\circ$. Data points are from \cite{Tsuji1998}. Black solid line corresponds to simulation with one standard deviation error band indicated.}
    \label{fig:Okj}
\end{figure}

Figure \ref{fig:Okj} shows that the simulated muon energy spectra follows well the data taken by Okayama's magnet spectrometer, the one standard deviation error band corresponds to the simulation statistical uncertainty. The ratio between the integrals of Monte Carlo and the experimental data is $1.15$, showing that the simulation describes correctly the data from the experiment. This value translates into a 15\% average difference between the simulation and the experiment, caused mainly by the generic atmospheric model used.

\subsection{\label{subsec:MuonBehaviour} Estimated Muon Flux at Bogota}

\begin{figure}[ht]
    \centering
    \includegraphics[width=0.8\textwidth]{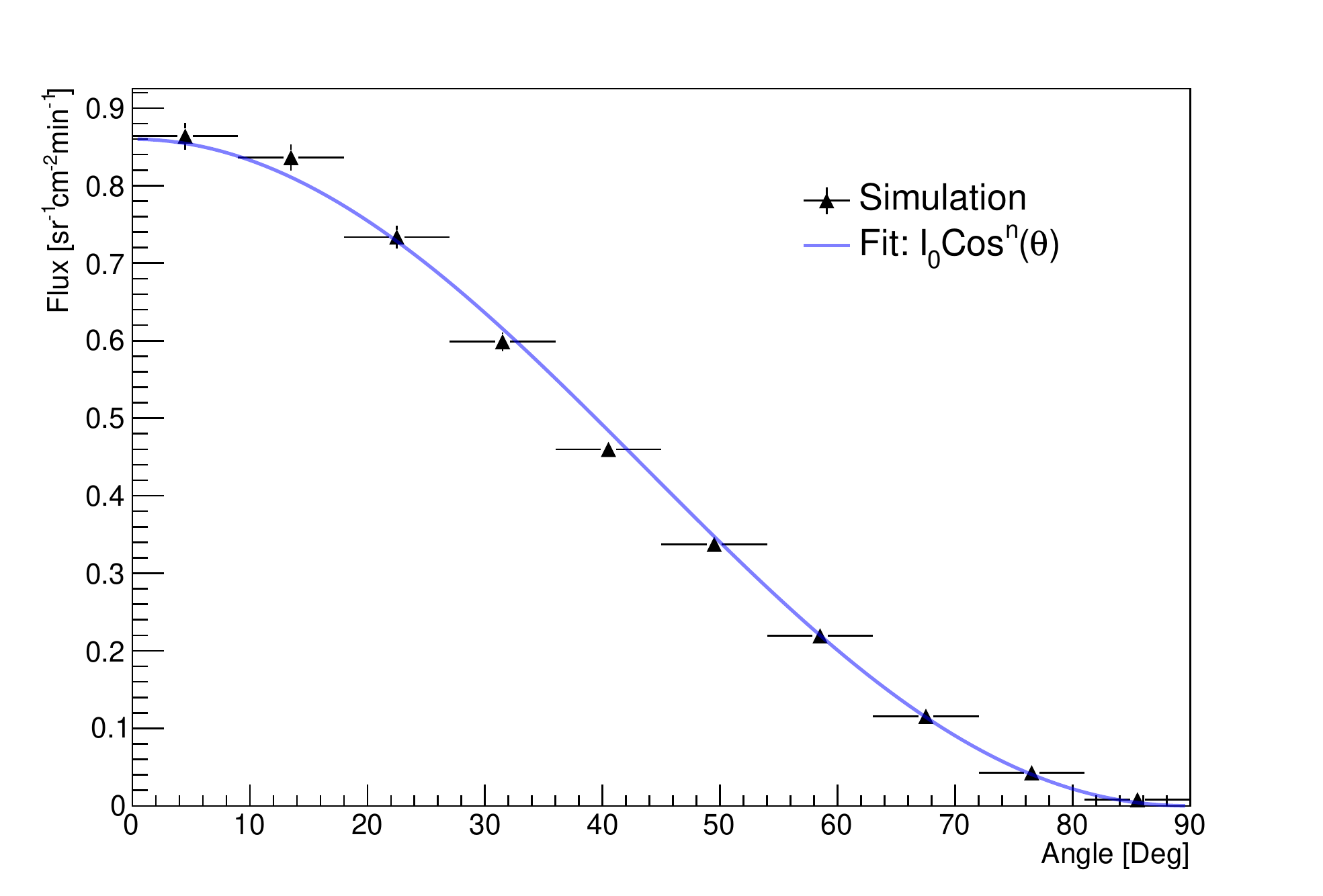}
    \caption{Simulated muon flux distribution (48 hours) as a function of zenith angle.}
    \label{fig:Theta}
\end{figure}

Provided the experimental verification of the CORSIKA simulation in sections \ref{subsec:FluxMeasurement} and \ref{subsec:FluxMeasurement2}, we proceed now to estimate the CR muon flux arriving in Bogota, Colombia. This estimate is quite interesting given that it provides insights about cosmic rays near the equatorial line (Bogota has a latitude of $4^\circ$). The simulation is carried out using the GPS coordinates presented in section \ref{sec:Monserrate}, the tropical atmospheric model provided by CORSIKA and magnetic field values of $B_x=26.9\;\mu T$, $B_z=14.3\;\mu T$. Figure \ref{fig:Theta} shows the muon flux intensity distribution, the best fit describing the data in the range $\theta=[0^{\circ}:70^{\circ}]$ corresponds to:
\begin{align}
    I(\theta)=  \frac{dN}{d\Omega dtdA} = &I_0\cos^n(\theta) \tagaddtext{[Muons \si{\per\centi\meter\squared\per\minute\per\steradian}]}\\
    I_0 = &0.85\pm 0.05\\
    n = & 2.11\pm0.03\\
    \chi^2/d.f. = & 1.08
\end{align}
For larger angles ($\theta\geq70^\circ$) the distribution shows a different behavior due to geomagnetic effects as indicated in \cite{cecchini2012atmospheric}.
Using this fit, the total vertical flux ($I_v$) for a horizontal detector at this location pointing directly to the sky is:
\begin{align}
    \label{eq:Ivertical}    
    I_v&=\int_{0}^{2\pi}\int_{0}^{\pi/2} I(\theta) d \Omega\\
    \begin{split}
        I_v &= 1.72 \pm0.09\tagaddtext{[Muons \si{\per\centi\meter\squared\per\minute}]}
        \label{eq:Iv1}
    \end{split}
\end{align}
The reported errors are only statistical. If a perfect cosine square distribution is assumed $ I(\theta)=I_0\cos^2(\theta)$ the total vertical flux can be calculated by integrating eq.(\ref{eq:Ivertical}) obtaining $I_v=2\pi I_0/3$. In our case the integral of eq.(\ref{eq:Ivertical}) is solved numerically using the parameter values given by the fit. 

Given that the validation of the CORSIKA simulation with experimental data showed an agreement at the 20\% level, we estimate a systematic uncertainty for the muon flux at Bogota at the same level. Therefore, the best estimate our studies provide for the vertical muon flux intensity at Bogota is:
\begin{equation}
    I_v = 1.72 \pm0.36\tagaddtext{[Muons \si{\per\centi\meter\squared\per\minute}]}
\end{equation}

As explained in sections \ref{subsec:FluxMeasurement} and \ref{subsec:FluxMeasurement2} the dominant source of error comes from the modeling of the atmospheric conditions.

\begin{figure}[ht]
    \centering
    \includegraphics[width=0.8\textwidth]{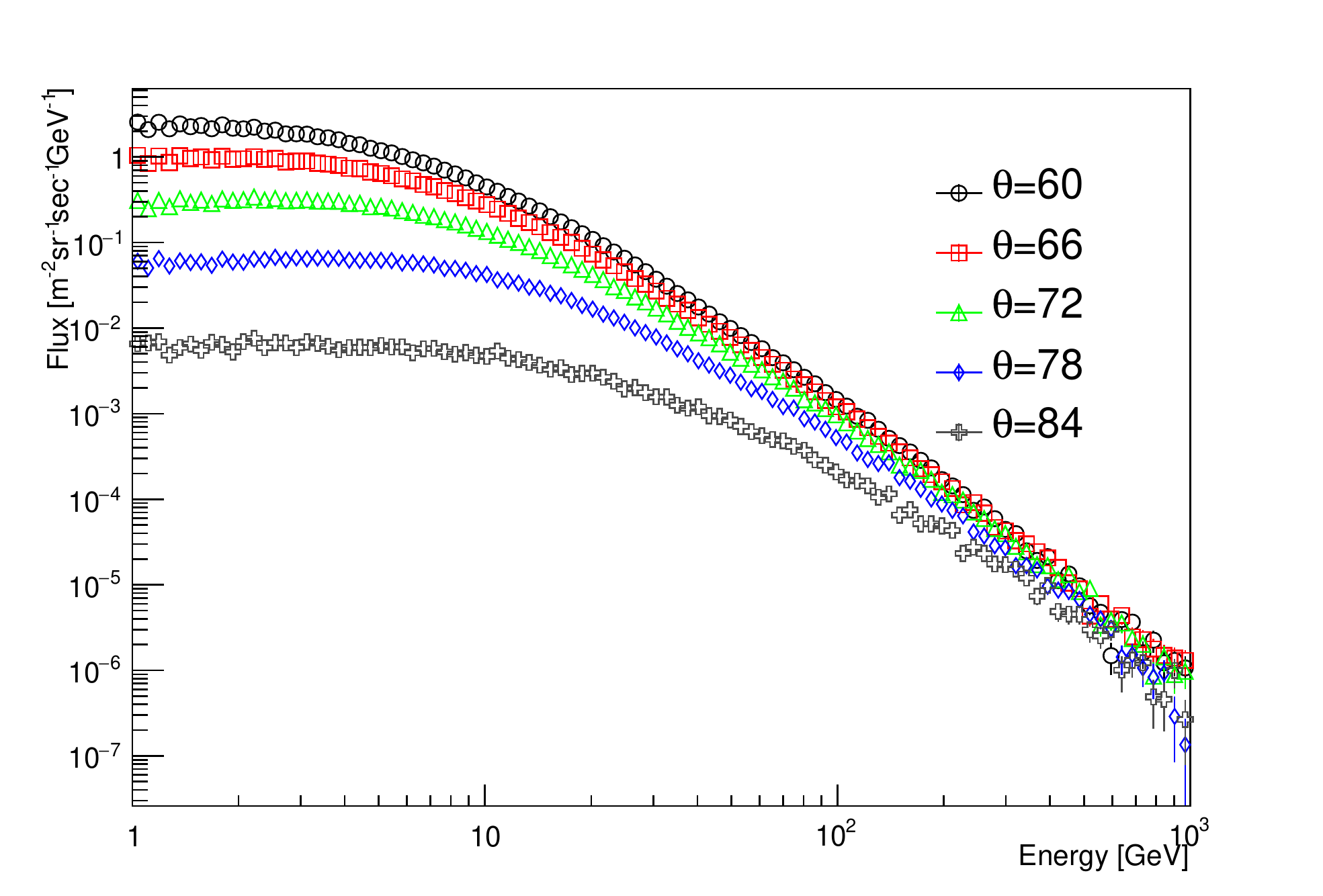}
    \caption{Energy spectra of muons ($\mu^{+}+\mu^{-}$), for 48h flux simulation. Five spectra are shown depending on different muon zenith angles of $\theta=60^\circ,66^\circ,72^\circ,78^\circ,84^\circ$ for the geographical location of Monserrate Hill.}
    \label{fig:Momentum}
\end{figure}

These results are compatible with the $I_v=1.4$ muons $\si{\per\centi\meter\squared\per\minute}$ reported by Morris in \cite{Morris2014} for an altitude of 2200 m above sea level which has a difference of 500 masl less than the simulation point used in our work.

The simulated muon energy spectrum is shown in figure \ref{fig:Momentum}. This spectrum is reported for five possible zenith angles of incidence $\theta=60^\circ,66^\circ,72^\circ,78^\circ,84^\circ$, which are selected taking into account the range of available incidence angles of muons into Monserrate Hill defining the region of interest. These results provide the expected behavior of the energy spectra, its relationship with the zenith angle of entry and are in accordance with the results of Allkofer \cite{allkofer1985cosmic} at sea level.

\begin{figure}[ht]
    \centering
    \includegraphics[width=0.8\textwidth]{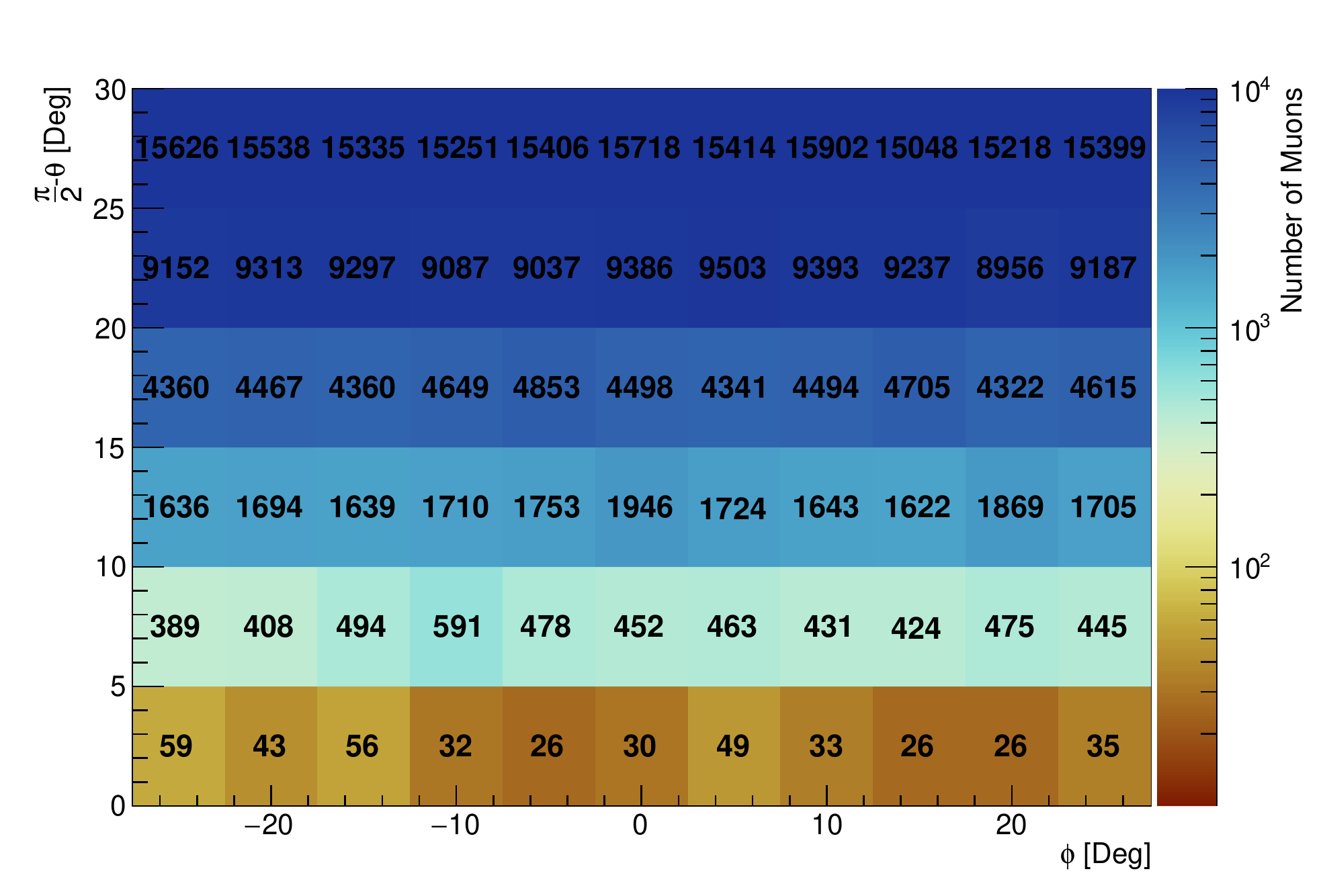}
    \caption{Simulation of 48 hours of muon flux arriving to a $1\;m\times 1\;m$ detector, located at our observation point, pointing towards open sky. }
    \label{fig:OSdistr}
\end{figure}

Figure \ref{fig:OSdistr} indicates the number of incident muons at open sky for ranges of zenith ($\theta$) and azimuth ($\phi$) angles of $\theta=[60^\circ:90^\circ]$ and $\phi=[-25^\circ:25^\circ]$, which define the coordinate system for Monserrate Hill. These results follow the trend reported in the literature \cite{Grieder2001}, showing a great number of counts for the lower part of the zenith range and an almost uniform distribution in the azimuth angle.

\section{\label{sec:MuonsMonserrate} Muons traversing Monserrate Hill}

In section \ref{sec:MuonFlux} we show that the CORSIKA simulations reproduce fairly well the cosmic muon spectral distributions and therefore, the simulations can be used to make estimations of muon flux transmission trough geological structures. Note that the systematical uncertainty quoted for the overall integrated flux is highly reduced for muon flux transmission studies because it depends on the ratio of transmitted muons over incident muons and any normalization effect gets washed out with the ratio. 
As seen at the bottom range of figure \ref{fig:OSdistr}, in average around 40 near horizontal muons are detected in 48 hours of incident cosmic ray flux. This number of muons is not enough to make any conclusions with significant accuracy. To generate enough statistics, a standalone Monte Carlo method is implemented for the simulation of muons crossing the mountain.

\subsection{\label{subsec:StandAloneMC} Standalone Monte Carlo Method}
The main purpose of this simulation is to generate a large number of muons ($\sim1\e{5}$) and calculate how many of those can cross the mountain and arrive at the detector.
The steps used by the Monte Carlo simulation are as follows:
\begin{enumerate}[I)]
    \item Generate a muon with coordinates $\mu(\theta,\phi)$ sampled from the cosine distribution (figure \ref{fig:Theta}) and a uniform distribution in $\phi=[-25^\circ:25^\circ]$.
    \item Given $\mu(\theta,\phi)$, a value of energy is randomly selected by using the distributions shown in figure \ref{fig:Momentum}, taking into account the value of the zenith angle $\theta$.
    \item Using an interpolation method for the muon energy loss in rock and the mean path inside the mountain, it is feasible to compute whether or not a generated muon can cross the expected path through the mountain (figure \ref{fig:LengthMonserrate}).
\end{enumerate}

Besides the simplicity of the method, it is still inefficient because a minimum of $10^{7}$ events must be generated to totally sample an energy distribution, but only the events with energies higher than $60$ $GeV$ (which account for less than 1\% of the total events generated) have a possibility to cross the mountain.

An improvement to the method is to limit the MC generation to a value of energy for which most of the generated muons have a high probability of traversing the mountain. Since the sampled limited distribution and the original one must be the same for their common energy range, we calculate a factor to properly normalize the generated events, by taking the ratio of the area on the corresponding distribution of figure \ref{fig:Momentum} for the limited energy range over the whole energy range:

\begin{equation}
    Flux_T = Flux_{E\geq E_0} \frac{A_{E\geq E_0}}{A_T}
    \label{eq:areaFactor}
\end{equation}

Where $E_0$ is the imposed energy limit and $A_{E\geq E_0}$, $A_T$ are the integrals of the limited and original energy distribution respectively, $Flux_T$ is the corrected normalized flux that emulates real conditions and $Flux_{E\geq E_0}$ corresponds to the flux estimation from the enriched sample of muons with a high probability of crossing the mountain. $Flux_{E\geq E_0}$ should be normalized by its integral to represent a probability distribution function needed for the proper event generation. For this study, the limits in energy are chosen to take into account the minimum energy needed to cross the mountain in each of the cells given by figure \ref{fig:EminMonserrate}.

From eq.(\ref{eq:areaFactor}) it is easy to see that the scaling factor is also sensitive to the integral of the original distribution, moreover, it has a dependency on the limits of that integral. The low part ($E<10$ $GeV$) is especially sensitive to geomagnetic effects that can prevent these particles to reach the detector \cite{lesparre2010geophysical}. Also, low momentum muons could be highly deflected from their original trajectory by multiple Coulomb scattering with the material of the mountain \cite{nishiyama2014experimental}. 

Several stochastic techniques can be used to calculate precisely these deflections \cite{kudryavtsev2009muon}\cite{niess2018backward}, but a simpler approach can be followed as used in \cite{borozdin2003surveillance}, by modeling the effect of Coulomb scattering as a Gaussian spread from the original muon trajectory \cite{tanabashi2018review}: 

\begin{equation}
    \frac{dN}{d\theta} = \frac{1}{\sqrt(2\pi)\theta_0}\exp{\frac{-\theta^2}{\theta_0^2}} 
    \label{eq:ScatteringCoulomb}
\end{equation}

Where $\theta$ is the incident angle of the muon and $\theta_0$ is the angular spread produced by the Coulomb interaction with atoms of the material the muon traverses, which depends on the radiation length ($l_0$) and the penetration distance $l$ \cite{tanabashi2018review} as:

\begin{equation}
    \theta_0=\frac{13.6MeV}{\beta c p}\sqrt{\frac{l}{l_0}}\big[1+0.038 \ln{(l/l_0)} \big]
    \label{eq:ScatteringCoulombDisp}
\end{equation}

\begin{figure}[ht]
    \centering
    \includegraphics[width=0.8\textwidth]{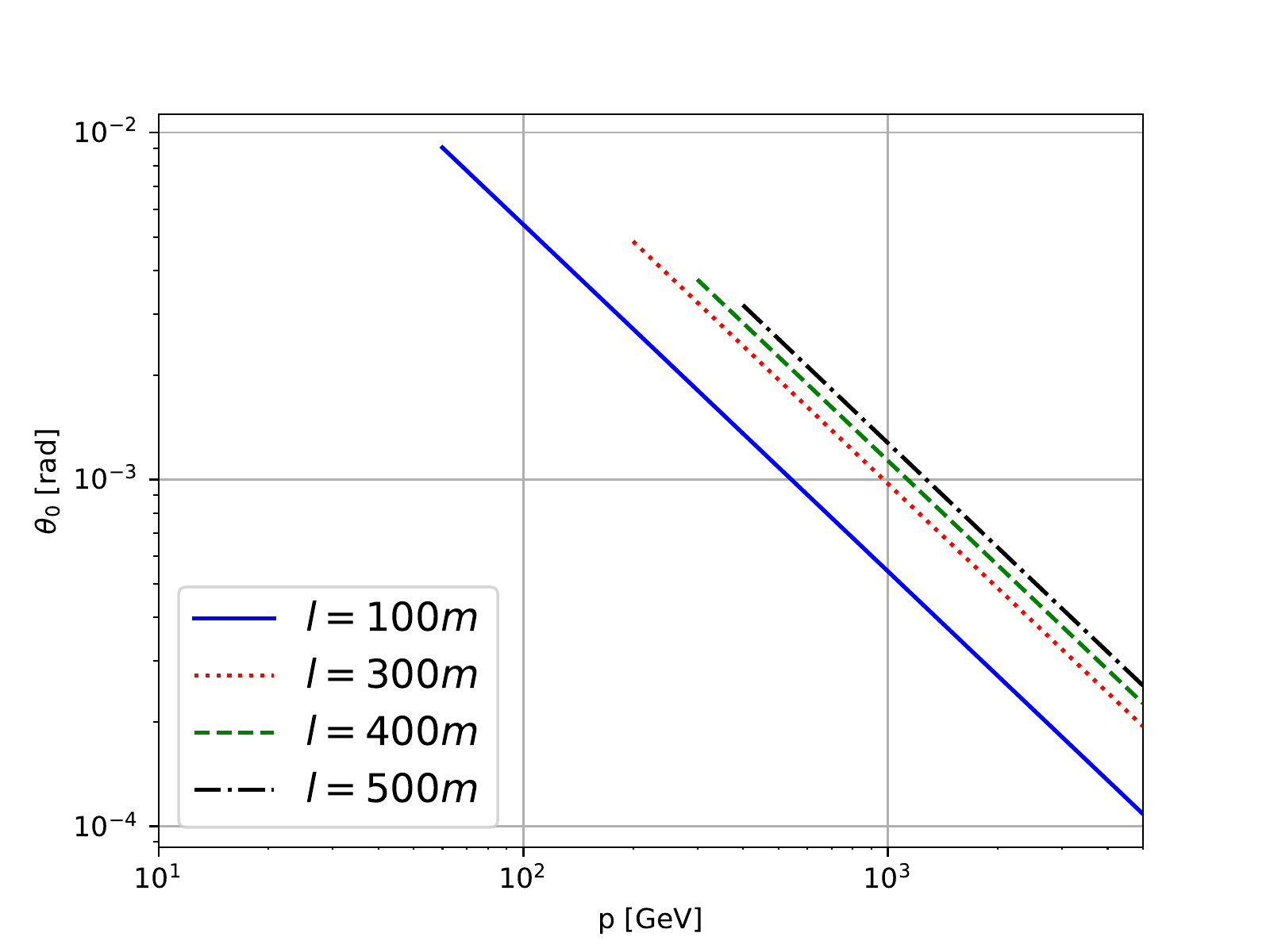}
    \caption{Muon angular spread as a function of incident momentum for different penetration lengths, $l$, in standard rock.}
    \label{fig:ScatteringDeflection}
\end{figure}

Figure \ref{fig:ScatteringDeflection} shows how the angular spread $\theta_0$ changes as a function of muon momentum and length traversed in standard rock. As indicated in figure \ref{fig:EminMonserrate}, the minimum energy to traverse a depth of $100\;m$ of standard rock is about $60\;GeV$, that corresponds to deviations lower than $10^{-2}rad$, providing smaller contributions to the systematic uncertainties as compared to other sources. 

\subsubsection{\label{subsubsec:InterpMethod} Interpolation Method for Energy Loss}

Using directly the Bethe-Bloch equation (eq.(\ref{eq:bethe})) can be quite cumbersome and computationally time-consuming due to the need of calculating all the parameters each time for every simulated muon. Also, this equation works well for an intermediate energy range (0.1 GeV--100 GeV) but for lower energies, when the particle's velocity becomes comparable to the electron velocities in the material the equation fails to provide an accurate description of the energy losses, and for higher energies radiation effects must be taken into account \cite{tanabashi2018review}.
Developing a program that uses the Bethe-Bloch equation taking into account all the limiting cases would lead to a whole new project, like the SRIM (Stopping and Range of Ions in Matter) \cite{ziegler2010srim}, GEANT4 \cite{Agostinelli:2002hh} and other similar programs \cite{mompart1996calculation}.

A simpler alternative is to use already existing tables of energy loss for standard rock as a function of incident muon momenta \cite{tanabashi2018review,GROOM2001}. Provided that we know the energy of the simulated muon, one can just use a linear interpolation from the existing tables to determine the average energy lost for a certain thickness, then recalculate the energy after traversing a thin material layer and keep iterating either until the muon traverses the whole length of the mountain or after it looses all of its energy and gets absorbed within the mountain. This procedure, in general, requires minimum computing time. Following the above procedure, one can just classify the muons into two categories: those absorbed by the mountain and those that cross it.

\begin{figure*}[ht]
    \centering
    \includegraphics[width=.8\textwidth]{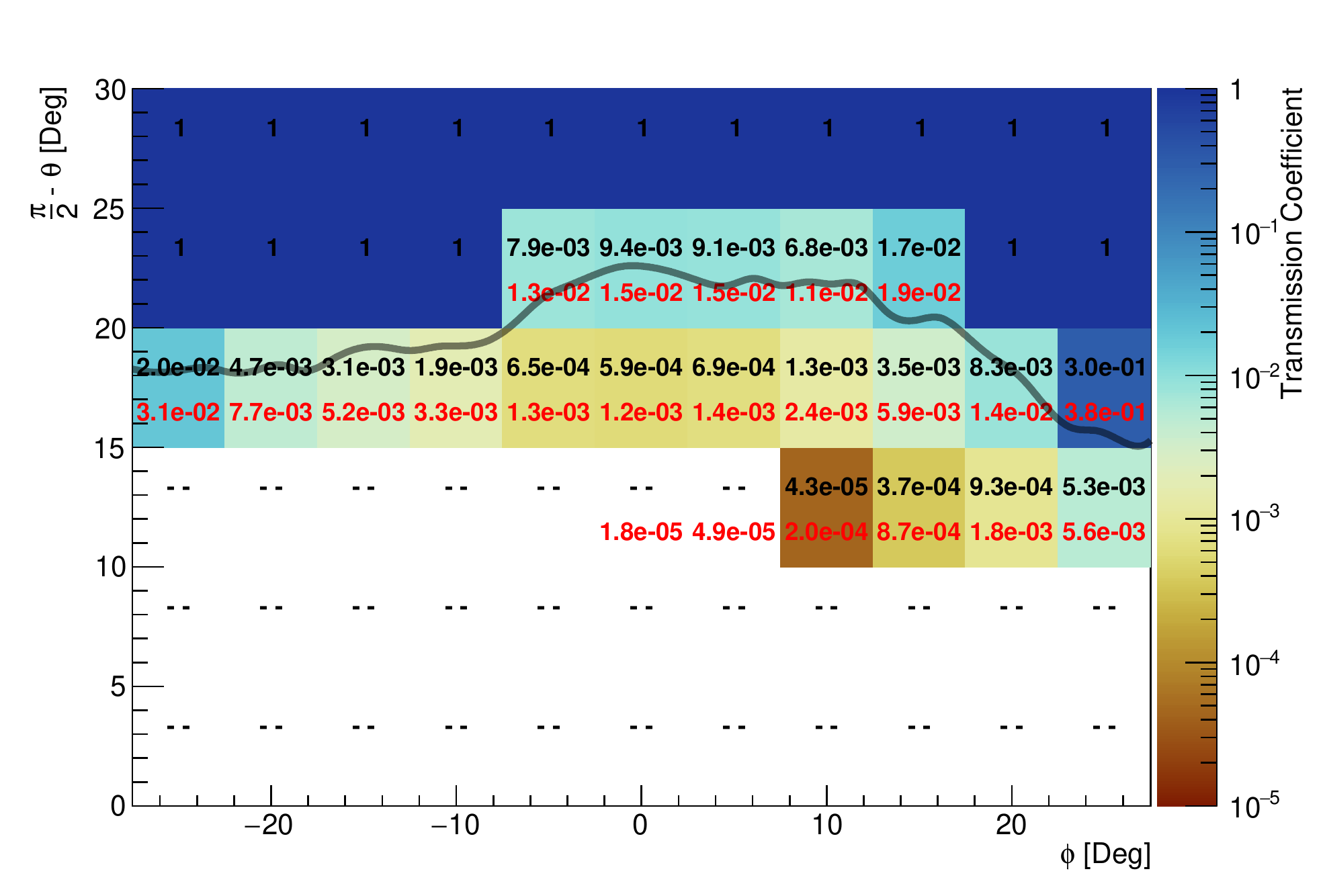}
    \caption{Muon transmission factor in Monserrate Hill. The black solid line corresponds to the mountain border as seen from the observation point in our university campus. Two values are displayed for each cell within the mountain, the top value corresponds to $\rho=2.65\;gcm^{-3}$ and the lower one to $\rho=2.11\;gcm^{-3}$. Cells with dashed lines imply that the expected transmission factor is lower than $10^{-5}$.}
    \label{fig:Trate}
\end{figure*}

\subsection{\label{subsec:TRate} Muon Transmission Factor}

With the results of the standalone MC process that corresponds to the number of muons that crossed the mountain, a transmission factor for a cell in the mountain is defined as the fraction of incident muons from open sky that cross a cell of the mountain with density $\rho$ and identified by angles $\theta$ and $\phi$ from the observation point. The fraction is corrected by any detector efficiency ($\varepsilon$):
\begin{equation}
    T(\theta,\phi,\rho) =\frac{1}{\varepsilon}\frac{N_{mtn}(\theta,\phi,\rho)}{N_{OS}(\theta,\phi)}
    \label{eq:trate}
\end{equation}

For our case $\varepsilon=1$, because of the assumption that the detector is able to identify every muon that arrives at it. $N_{OS}(\theta,\phi)$ and $N_{mtn}(\theta,\phi,\rho)$ correspond to the number of muons that arrive at a given cell defined by ($\theta,\phi$) and the ones that cross a similar cell with density $\rho$, respectively, both numbers measured within the same time interval.

Figure \ref{fig:Trate} shows the estimated muon transmission factor for two different rock densities ($\rho=2.65\;gcm^{-3}$ and $\rho=2.11\;gcm^{-3}$). The lower density approaches to the density values for sandstone found in Monserrate \cite{lobo1992geologia}. Note that the attenuation factor for a specific cell is just the transmission factor for that cell subtracted from unity.

We just make a quick comparison of our results with muon transmission factor measurements from other authors as follows: for Mt Echia the transmission factor for a distance of 70 m and a density of $\rho=1.4\;gcm^{-3}$ is about $T_{Echia}\approx6.5\e{-2}$ \cite{saracino2017imaging}, and we are obtaining $T_{Monserrate}\approx1.7\e{-2}$ with a density of $2.65\;gcm^{-3}$ and a distance of about 100 m. Our result for Monserrate Hill is about $3.8$ times lower than the transmission factor measured for Mt Echia, this difference is primarly due to the distinct values of density and the path length traversed by muons. Tanaka et al \cite{tanaka2007high} measured the change in the number of muons through Mt. Asama, reporting muon transmission factors between $8.5\e{-5}-3.0\e{-3}$ for densities between $\rho=2.36\;gcm^{-3}$ and $2.52\;gcm^{-3}$. Our estimated transmission factors are on the same range of values. 

\begin{figure}[ht]
\centering
  \includegraphics[width=0.8\textwidth]{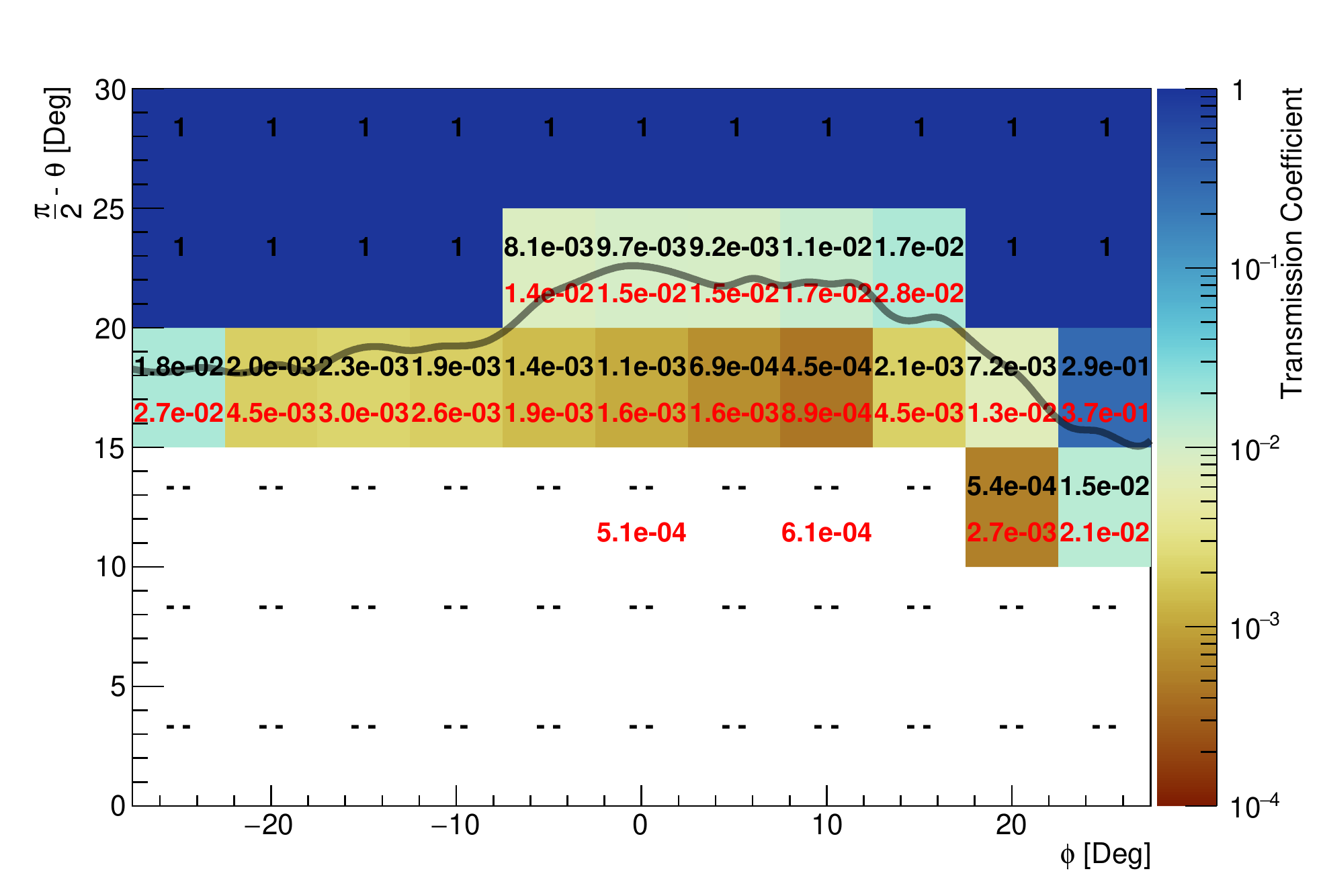}
\caption{Muon's transmission factor for Monserrate Hill with densities of $\rho=2.65\;gcm^{-3}$ (upper value) and $\rho=2.11\;gcm^{-3}$ (lower value)  and 48 hours of CORSIKA simulation, calculated with the direct method explained in section \ref{subsec:Trate2}. Cells with dashed lines imply that the expected transmission is lower than $10^{-4}$}
\label{fig:MtnCSK}
\end{figure}

\subsection{\label{subsec:Trate2} Transmission Factor with CORSIKA}
An alternative and more direct method for the calculation of the transmission factor is to take the muons simulated directly by CORSIKA (figure \ref{fig:OSdistr}), then make them cross the mountain and take the ratio as in eq.(\ref{eq:trate}). However, this method has a direct problem involving the computing time of the events, for example, 2 hours of shower simulations take around 4 hours of computing time in the computing system used for our studies. For high statistics the standalone MC method, described in section \ref{subsec:StandAloneMC}, is the best alternative.\\

To validate our standalone MC results with CORSIKA results, a simulation of 48 hours of shower events and a mountain composed of standard rock with uniform density was performed. The number of muons detected at open sky is reported in figure \ref{fig:OSdistr}, the transmission factors of muons that cross the mountain are shown in figure \ref{fig:MtnCSK}. These results are in good agreement with the ones obtained using the standalone MC method.

\begin{figure*}[ht]
    \centering
    \includegraphics[width=0.8\textwidth]{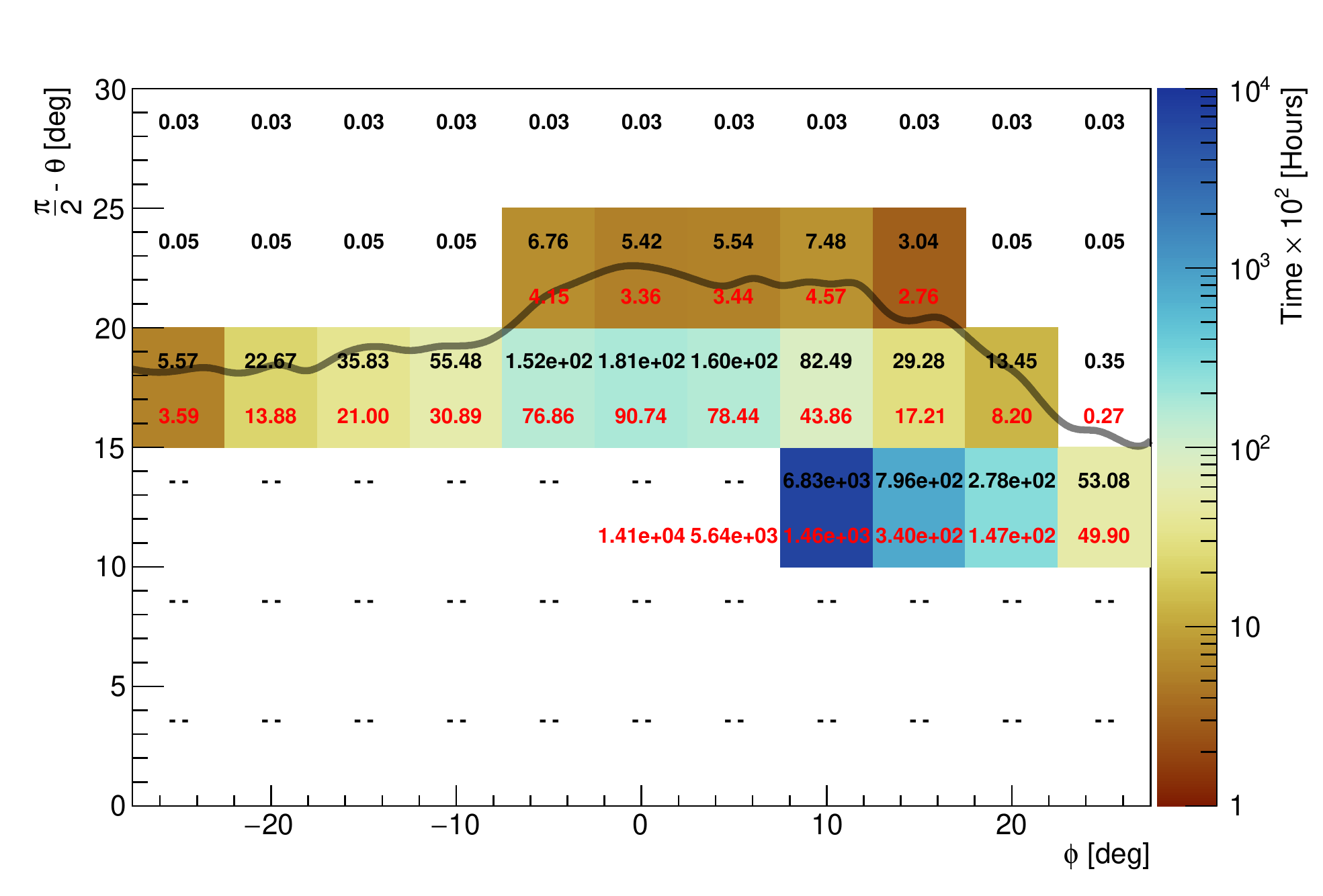}
    \caption{Time estimate to detect 1000 muons in each cell of Monserrate Hill. Upper (lower) value corresponds to a density of $2.65\;gcm^{-3}$ ($2.11\;gcm^{-3}$). Cells with dashed lines imply that the estimate time to have 1000 muons traversing the cell is greater than $10^7$ hours. For the cells covering the border of the mountain, we are only considering the muons going trough the mountain and ignoring the open sky muons that also traverse that cells.}
    \label{fig:TimePred}
\end{figure*}

\subsection{\label{subsec:TimePred} Time Prediction}

As a final calculation, using the transmission factor and the behavior of muons at open sky from the standalone MC, for a simulation of flux during a time $\Delta t_{sim}$ (in hour case 24 hours), it is possible to estimate the time that would take for an experiment to collect certain number of muon, $N_\mu$, traversing each cell of the mountain. Since the transmitted muon flux for a specific cell integrated over time  has to be equal to the desired number of muons, over area and solid angle, that we want to gather from that cell, we can extract the time needed for data taking as:
\begin{equation}
  \Delta t = \frac{N_{\mu} \;\Delta t_{sim}}{T(\theta,\phi,\rho)*N_{OS(\theta,\phi)}}
  \label{eq:TimePred}
\end{equation}

In our case, we make the time estimate to collect 1000 muons which have a statistical uncertainty near 3\%.

Figure \ref{fig:TimePred} indicates the estimated time in hours needed to obtain 1000 muons traversing different cells of Monserrate Hill for two mountain densities. These data might be useful to constraint the time requirements of an experiment, for example, the measured time is around 3-5 hours for open sky muons but for the ones crossing Monserrate Hill the minimum time is about 300 hours. It is worth to notice that when a cell has a prediction time of "--" that means that none of the $1\e{5}$ muons produced by the MC process have crossed the mountain.

From these results, a better approach to the detector system can be suggested, that is to improve the solid angle covered by the detector to gather data from several of these cells simultaneausly with enough spatial resolution to identify the cell crossed by each detected muon. Also, the most feasible region to study in Monserrate Hill is the top region ($\theta=[65^\circ:70^\circ],\phi=[-5^\circ:20^\circ]$).

\section{\label{sec:Conclusions} Conclusions}
The data analysis techniques developed in this work grant a simple and direct way to simulate a muon tomography of a mountain. Our results are backed by replicating experimental data in different locations around the world. 
Improvements in the modeling of the atmosphere, geomagnetic conditions and material composition of the mountain could be implemented for more accurate estimates but may require more complex work. We just prove that with a simple approach, fast estimates of muon transmission can be obtained along with time estimates for data acquisition according to the desired accuracy.
If a 3\% error is desired in the measurement of the number of muons that cross approximately 100 m of standard rock, the experiment would need to gather data for over 12 days, reducing this time can be achieved by increasing the solid angle covered by the detector system and by selecting a feasible region to scan. For a mountain composed of standard rock, the feasible region should have a depth ranging from 100 m--1000 m to keep the data taking period within a one-year time frame.
The results obtained can be extrapolated to different locations with particular characteristics in an easy way, providing constraints and serving as a guide for in-situ experiments in muon tomography.

\acknowledgments
We acknowledge financial support from the Physics Department and the Faculty of Science of Universidad de los Andes, Colombia, under project number P18.160322.001-15FISI01.
\bibliographystyle{JHEP}
\bibliography{main}

\end{document}